\documentclass{article}
\usepackage{times}
\usepackage{amsfonts}
\usepackage{graphicx}
\usepackage[pdfmark]{hyperref}
\begin{document}
\noindent
{\Large ON THE NONCOMMUTATIVE EIKONAL}
\vskip1cm
\noindent
{\bf J.M. Isidro}${}^{1,2,a}$, {\bf P. Fern\'andez de C\'ordoba}${}^{1,b}$,  {\bf J.M. Rivera--Rebolledo}$^{3, c}$\\
and {\bf J.L.G. Santander}${}^{4,d}$\\
${}^{1}$Instituto Universitario de Matem\'atica Pura y Aplicada,\\ Universidad Polit\'ecnica de Valencia, Valencia 46022, Spain\\
${}^{2}$Max--Planck--Institut f\"ur Gravitationsphysik, Albert--Einstein--Institut,\\ D--14476 Golm, Germany\\
${}^{3}$Departamento de F\'{\i}sica, U. P. Adolfo L\'opez Mateos, 07738 Lindavista,\\
M\'exico D. F., Mexico\\
${}^{4}$Departamento de Ciencias Experimentales y Matem\'aticas,\\ Universidad Cat\'olica de Valencia, Valencia 46002, Spain\\
${}^{a}${\tt joissan@mat.upv.es}, ${}^{b}${\tt pfernandez@mat.upv.es}, \\
${}^{c}${\tt jrivera@esfm.ipn.mx}, ${}^{d}${\tt jlgonzalez@mat.upv.es} \\
%\vskip.5cm
%\noindent
%\today
\vskip.5cm
\noindent
{\bf Abstract} We study the eikonal approximation to quantum mechanics on the Moyal plane. Instead of using a star product, the analysis is carried out in terms of operator--valued wavefunctions depending on noncommuting, operator--valued coordinates.

\section{Introduction}\label{einfuehrung}

That spacetime must stop being a continuum once sufficiently high energies are reached is by now an old notion. Already in the 1930's, Heisenberg replaced a continuum spacetime with a lattice, in order to tame the divergences of quantum field theory. This lattice broke Lorentz invariance, which later models \cite{SNYDER} succeeded in preserving. Attempts to quantise gravity also lead to the introduction of a fundamental length scale. This fundamental length scale, beyond which the concept of distance becomes meaningless, is called the Planck length. Replacing a continuum with some kind of discrete, or quantised, manifold, leads naturally to the conclusion that coordinates must be operator--valued quantities.

The purpose of this paper is to analyse the eikonal (or semiclassical) approximation to nonrelativistic quantum mechanics on the Moyal plane $\mathbb{R}_{\theta}^2$, the latter being coordinatised by operators $X,Y$ satisfying the Heisenberg algebra $[X,Y]=XY-YX={\rm i}\theta{\bf 1}$, with $\theta>0$. In particular we will need to write down the Hamilton--Jacobi equation, and its close cousin the Schroedinger equation, on the noncommutative plane. Two steps are involved here:\\
{\it i)} defining a classical mechanics on the noncommutative plane $\mathbb{R}_{\theta}^2$;\\
{\it ii) } quantising the classical mechanics so defined.\\
On an energy scale, quantisation ($\hbar> 0$) sets in well before noncommutativity ($\theta> 0$). In other words, in the real world one expects mechanics on a noncommutative space to be automatically quantum, rather than classical. We will see that, in fact, steps  {\it i)} and {\it ii)} above are inextricably linked. However, if only methodologically, we will consider the two steps above separately.

There are several alternative, though basically equivalent, approaches to physics on noncommutative spaces. One approach, by far the most widespread, uses number--valued coordinates and momenta (and functions thereof), while replacing the commutative pointwise product of functions with a noncommutative star product \cite{SARDA}. Another approach, little developed so far, uses operator--valued coordinates and momenta already from the start.  Coordinates and momenta are now multiplied as matrices. In particular, operator--valued coordinates satisfy certain nontrivial commutation relations. In this paper we will further develop this second approach: we want our wavefunctions $\Psi$ to be functions of the position operators $X,Y$, the latter satisfying $[X,Y]={\rm i}\theta{\bf 1}$. This will imply that the wavefunction $\Psi$ itself will become an operator. This property is reminiscent of second--quantised theories. In fact it has been argued \cite{NAIR} that (yet another) approach to noncommutative quantum mechanics is provided by the 1--particle sector of noncommutative quantum field theory; then the noncommutative wavefunction, a c--number object, arises as a matrix element of a field operator\footnote{Of course, the previous statement also applies to theories on commutative spacetimes.}. However we prefer to stay within the framework of a finite number of degrees of freedom, and to try and construct wavefunctions that are operator--valued from the start, without resorting to field theory. After all, at least on commutative spaces, it is perfectly possible to formulate the quantum mechanics of a finite number of degrees of freedom, in a manner that is totally independent of quantum field theory, {\it i.e.}, without embedding quantum mechanics into an infinite number of degrees of freedom. We will see that noncommutative spaces also share this possibility. One surprising outcome of our approach will be that the mechanical action itself (the solution to the Hamilton--Jacobi equation) will also become an operator, without second--quantising the theory.

One can turn our argument around and analyse to what extent quantum mechanics, especially {\it emergent}\/ quantum mechanics \cite{ELZE} and also {\it emergent}\/ gravity \cite{SEIBERG, PADDY, SINGH}, imply a granularity of spacetime. A comprehensive exposition of emergent physics, with extensive references, is given in the nice book \cite{CARROLL0}. The existing literature on noncommutative theories (deformation quantisation, quantum mechanics, field theory, string theory, gravity) is too vast to quote here, but we would like to mention the general refs. \cite{MADORE, SZABO, INDIO, MASINDIOS}. Geometric treatments of quantum mechanics have also been studied in depth; for a sample see, {\it e.g.}, \cite{SARDA, CARROLL1, GOSSON, NOI, KOCH}.

Approaches to quantum theory that are primarily based on Hamilton--Jacobi equations (and generalisations thereof) are well known; we will just mention \cite{MATONE} and the many references therein. For later use we recall that classical Hamilton--Jacobi theory on the commutative configuration space $\mathbb{R}^2$ coordinatised by $x,y$ can be (extremely succintly) summarised by the equations
\begin{equation}
{\cal S}=-Et+\int p_x{\rm d}x+p_y{\rm d}y, \qquad p_x=\frac{\partial {\cal S}}{\partial x},\qquad p_y=\frac{\partial {\cal S}}{\partial y}
\label{aktion}
\end{equation}
and
\begin{equation}
\frac{\partial {\cal S}}{\partial t}+\frac{1}{2m}\left[\left(\frac{\partial {\cal S}}{\partial x}\right)^2+\left(\frac{\partial {\cal S}}{\partial y}\right)^2\right]+{\cal U}(x,y)=0,
\label{kaspos}
\end{equation}
where ${\cal U}(x,y)$ is a potential function. We use the notation ${\cal S}$ for the {\it classical}\/ action integral on the commutative plane $\mathbb{R}^2$, in order to distinguish  it from the {\it operator--valued action}\/ $S$ to be introduced presently on the Moyal plane $\mathbb{R}^2_{\theta}$. The same notational convention applies to the classical potential function ${\cal U}$ and to its operator--valued analogue $U$, to be defined later.

This paper is organised as follows. Section \ref{nonpo} presents the noncommutative algebra of position and momentum operators that our construction is based upon. This algebra can be unitarily represented in a number of different ways. For reasons that will become clear presently, our favourite representation is given in terms of {\it noncommutative oscillator modes}\/. The latter can be thought of as harmonic oscillators on (an auxiliary copy of) the Moyal plane; we provide an explicit construction of these noncommutative oscillator modes, in some detail. Once position and momentum operators $X,Y,P_X,P_Y$ are defined in terms of these modes, we need to define a mechanical action $S$ depending on $X,Y,P_X,P_Y$ and such that properties as close as possible to those satisfied by its commutative counterpart (\ref{aktion}), (\ref{kaspos}) continue to hold true. This is done in section \ref{hjmp}. By now we have an object $S$ that plays the role of the classical mechanical action ${\cal S}$. However $S$ is operator--valued, because the position and momentum variables it depends on are themselves operators. The next step, at least in a semiclassical analysis, is to consider the exponential of ($i$ times) $S$, and to derive the equation satisfied by the latter, the Schroedinger equation on noncommutative space. This is done in section \ref{plkdmr}. Despite the numerous formal analogies with quantum mechanics on the commutative  plane, there are some substantial differences that are pointed out along the way. Finally section \ref{diskku} presents some concluding remarks concerning:\\
{\it i)} the role of the Bopp shift and the nonequivalent Poisson structures that it relates;\\
{\it ii)} the commutative limit of our model;\\
{\it iii)} the resolution of some apparent clashes with some classical theorems of Wigner, and of Stone and von Neumann;\\
{\it iv)} some speculations about a classical/quantum duality in noncommutative theories.\\

\section{The noncommutative Poisson--Heisenberg algebra}\label{nonpo}

\subsection{The commutator algebra}\label{ssuno}

The noncommutative plane $\mathbb{R}_{\theta}^2$ is defined as the algebra of functions of two generators $X,Y$ satisfying the commutator $[X,Y]={\rm i}\theta{\bf 1}$, with $\theta>0$. We regard $\mathbb{R}_{\theta}^2$ as a two--dimensional configuration space endowed with noncommuting coordinates $X,Y$. On the corresponding noncommutative phase space $\mathbb{R}_{\theta, \hbar}^4$ we have the operators $X$, $Y$, $P_X$, $P_Y$ satisfying a commutator algebra that we postulate to be
\begin{equation}
[X,Y]={\rm i}\theta {\bf 1},\quad [X,P_X]=[Y,P_Y]={\rm i}\hbar{\bf 1},\quad [P_X,P_Y]=[X,P_Y]=[Y,P_X]=0,
\label{possonalg}
\end{equation}
We will call the set of eqns. (\ref{possonalg}) the  2--dimensional, {\it noncommutative Poisson--Heisenberg algebra}\/.  The time variable $t$ will be taken to commute with all generators $X,Y,P_X,P_Y$.

It has been known for long that the {\it Bopp shift}
\begin{equation}
Y\mapsto Y-\frac{{\theta}}{\hbar}P_X
\label{vop}
\end{equation}
reduces the noncommutative Poisson--Heisenberg algebra (\ref{possonalg}) to the usual Poisson--Heisenberg algebra in two commuting space dimensions. This notwithstanding, it is instructive to work with the algebra (\ref{possonalg}). This is so because one can think of (\ref{possonalg}) as being the commutator algebra of quantum mechanics with {\it two}\/ deformation parameters---one quantum of area $\theta$, one quantum of action $\hbar$. Standard quantum mechanics contains only the quantum of area on phase space, $\hbar$; noncommutative quantum mechanics adds a quantum of area on configuration space, $\theta$. In the presence of the two quanta $\hbar$ and $\theta$, and given a particle of mass $m$, the quantity $\hbar^2/(m\theta)$ has the dimensions of energy. We will see that the quantity $\hbar^2/(m\theta)$ plays an important role in what follows.

As usual we define the adjoint action of operator $A$ on operator $B$  by
\begin{equation}
{\rm ad}_A(B)=[A,B].
\label{adjk}
\end{equation}
The adjoint action ${\rm ad}_A(B)$ behaves formally as a derivative: it is linear and satisfies the Leibniz rule
\begin{equation}
{\rm ad}_A\left(BC\right)={\rm ad}_A\left(B\right)\,C+B\,{\rm ad}_A\left(C\right).
\label{leib}
\end{equation}
We also have the Jacobi identity
\begin{equation}
\left[{\rm ad}_A, {\rm ad}_B\right]={\rm ad}_{[A,B]},
\label{homor}
\end{equation}
which expresses a generalisation of the integrability property $\partial^2 f/\partial x\partial y=\partial^2 f/\partial y\partial x$ valid for derivatives of functions $f(x,y)$. Replacing phase--space derivatives with adjoint actions will be an essential tool in our approach to noncommutative quantum mechanics.

\subsection{Commutative oscillator modes}\label{com}

We will first construct a Hilbert--space representation for the commutator algebra (\ref{possonalg}), in terms of {\it commutative}\/ oscillator modes. This is of course trivial, but it will serve as a warmup exercise for the construction in terms of {\it noncommutative}\/ oscillator modes. Consider the usual harmonic oscillator eigenstates $\phi_n$ in 1 dimension,  where $n\in\mathbb{N}$. The space spanned by the $\phi_n$ is ${\ell}^2$, the Hilbert space of complex, square--summable sequences.
In two commuting dimensions $x,y$ we have the eigenstates $\phi_{nm}(x,y)=\phi_{n}(x)\phi_{m}(y)$. The latter form an orthonormal basis for the Hilbert space ${\ell}^2\times {\ell}^2$. Position and momentum operators $X',Y',P'_X,P'_Y$ can be defined on the space ${\ell}^2\times {\ell}^2$ as usual \cite{LANDAU}: acting on the first index,
\begin{equation}
X'\phi_{nm}:=\sqrt{\frac{\theta}{2}}\left(\sqrt{n+1}\,\phi_{n+1,m}+\sqrt{n}\,\phi_{n-1,m}\right),
\label{unoequis}
\end{equation}
\begin{equation}
P'_X\phi_{nm}:=\frac{{\rm i}\hbar}{\sqrt{2\theta}}\left(\sqrt{n+1}\,\phi_{n+1,m}-\sqrt{n}\,\phi_{n-1,m}\right).
\label{unopequis}
\end{equation}
{}For the second index we define the action of $Y',P'_Y$ similarly, with the sole difference that the (reverse) Bopp shift (\ref{vop}) must be taken into account:
\begin{equation}
Y'\phi_{nm}:=\sqrt{\frac{\theta}{2}}\left(\sqrt{m+1}\,\phi_{n,m+1}+\sqrt{m}\,\phi_{n,m-1}\right)+\frac{\theta}{\hbar}P'_X\phi_{nm},
\label{unoygriega}
\end{equation}
\begin{equation}
P'_Y\phi_{nm}:=\frac{{\rm i}\hbar}{\sqrt{2\theta}}\left(\sqrt{m+1}\,\phi_{n,m+1}-\sqrt{m}\,\phi_{n,m-1}\right).
\label{unopygriega}
\end{equation}
One verifies that the operators $X',Y',P'_X,P'_Y$ indeed satisfy the algebra (\ref{possonalg}).  We have denoted these operators with a prime because this representation is unsatisfactory for our purposes. Indeed, there is nothing noncommutative about the eigenstates $\phi_{nm}$: they are simply those of the harmonic oscillator on the commutative plane $\mathbb{R}^2$, noncommutativity being implemented in the algebra by means of the (inverse) Bopp shift. Instead one would like to have a representation space spanned by eigenstates $\psi_{nm}$ of the harmonic oscillator on the noncommutative plane $\mathbb{R}^2_{\theta}$. This will be done explicitly in section \ref{sstres}.

\subsection{Interlude}\label{ssdos}

Before moving on to noncommutative oscillator modes we need to recall some elementary facts \cite{THIRRING}. Consider the space $F$ of all entire functions $f:\mathbb{C}\rightarrow\mathbb{C}$ such that
\begin{equation}
f(z)=\sum_{n=0}^{\infty}\frac{c_n}{\sqrt{n!}}z^n, \qquad \sum_{n=0}^{\infty}\vert c_n\vert^2<\infty.
\label{bese}
\end{equation}
This space is Hilbert with respect to the scalar product
\begin{equation}
\langle f\vert \tilde f\rangle:=\frac{1}{2\pi{\rm i}}\int {\rm d}z^*\wedge {\rm d}z\, {f^*(z)}\tilde f(z){\rm e}^{-\vert z\vert^2},
\label{presc}
\end{equation}
where the asterisk denotes complex conjugation, and the integral extends over all $\mathbb{R}^2$ with $z=(x+{\rm i}y)/\sqrt{2}$. An orthonormal basis is given by the set of all complex monomials
\begin{equation}
f_n(z):=\frac{z^n}{\sqrt{n!}}, \qquad n\in\mathbb{N}.
\label{momo}
\end{equation}
The space $F$ is called {\it Bargman--Segal space}\/. The $f_n$ are in 1--to--1 correspondence with the harmonic oscillator eigenstates $\phi_n$ of section \ref{com}.

Next consider the following variant of Bargman--Segal space. Let us consider functions  $g:\mathbb{R}\rightarrow\mathbb{C}$  such that
\begin{equation}
g(x)=\sum_{n=0}^{\infty}\frac{c_n}{\sqrt{n!}}x^n, \qquad \sum_{n=0}^{\infty}\vert c_n\vert^2<\infty,
\label{besereal}
\end{equation}
the $c_n$ being complex coefficients. Here our functions $g$ are complex--valued analytic functions of one {\it real}\/ variable $x$. Call $G$ the space of all functions satisfying (\ref{besereal}). A basis for $G$ is given by the set of all real monomials
\begin{equation}
g_n(x):=\frac{x^n}{\sqrt{n!}}, \qquad n\in\mathbb{N}.
\label{basereal}
\end{equation}
We can define a scalar product on $G$ by declaring these monomials to be orthonormal,
\begin{equation}
\langle g_n\vert g_m\rangle :=\delta_{nm}, \qquad n,m\in\mathbb{N},
\label{toro}
\end{equation}
and extending the above to all elements of $G$ by complex linearity. This scalar product makes $G$ a complex Hilbert space. The difference with respect to Bargman--Segal space $F$ is that, the functions $g\in G$ depending on the real variable $x$ instead of the complex variable $z$, the scalar product on $G$ is no longer given by (\ref{presc}), nor by its real analogue. Indeed, given any two $g,\tilde g\in G$, the analogue of (\ref{presc}) for $G$ would be the integral
\begin{equation}
\int_{-\infty}^{\infty}{\rm d}x\,g^*(x)\tilde g(x){\rm e}^{- x^2}.
\label{noupre}
\end{equation}
Although this integral does define a scalar product on $G$, this scalar product does not make the basis (\ref{basereal}) orthogonal, as one readily verifies. Therefore  one, and only one, of the following properties can be satisfied:\\
{\it i)} the space $G$ is Hilbert with respect to the scalar product (\ref{noupre}), but the monomial basis (\ref{basereal}) is not orthogonal with respect to it;\\
{\it ii)} the space $G$ is Hilbert with respect to the scalar product (\ref{toro}), and the monomial basis (\ref{basereal}) is indeed orthonormal with respect to it, but this scalar product is not given by the integral (\ref{noupre}).\\
This being the case, we settle in favour of condition {\it ii)} above as our choice for the Hilbert space $G$.

{}Finally, the construction given by eqns.  (\ref{besereal})--(\ref{toro}) can be straightforwardly extended to complex--valued, analytic functions of {\it two}\/ real variables $x,y$. This will be used next.

\subsection{Noncommutative oscillator modes}\label{sstres}

Next we construct a unitary, Hilbert--space representation for the algebra (\ref{possonalg}), in terms of noncommutative oscillator modes.  It will be based on the Hilbert space, just mentioned in section \ref{ssdos}, of complex--valued, analytic functions of two real variables---but with {\it noncommuting, selfadjoint operators}\/ replacing the real variables.

Consider first an auxiliary copy ${\cal H}$ of the Heisenberg algebra, spanned by operators $V,W,{\bf 1}$ satisfying $[V,W]={\rm i}\theta {\bf 1}$, where both $V$ and $W$ have dimensions of length. The algebra ${\cal H}$ is realised in the standard way: $V$ acts on auxiliary wavefunctions $h(v)$ by multiplication, $Vh(v)=vh(v)$, and $W$ acts by differentiation, $Wh(v)=-{\rm i}\theta{\rm d}h/{\rm d}v$.  That the dimension of $\theta$ is length squared, rather than that of an action, should not bother us, since ${\cal H}$ is an auxiliary construct. The corresponding Hilbert space of the wavefunctions $h(v)$, also termed auxiliary, is $L^2(\mathbb{R}, {\rm d}v)$. This Hilbert space, however, is {\it not}\/ the carrier space of the unitary representation of the algebra (\ref{possonalg}) that we are looking for. To reiterate, the algebra $[V,W]={\rm i}\theta {\bf 1}$ just introduced, although  isomorphic to the subalgebra $[X,Y]={\rm i}\theta {\bf 1}$ contained in (\ref{possonalg}), acts on the auxiliary space $L^2(\mathbb{R}, {\rm d}v)$, while the space on which the algebra $[X,Y]={\rm i}\theta {\bf 1}$ will act is about to be defined below.

Next let $U({\cal H})$ denote the universal enveloping algebra of ${\cal H}$. By definition, $U({\cal H})$ is the algebra of  polynomials in the operators $V,W,{\bf 1}$, of arbitrarily high degree, with $V$ and $W$ satisfying $[V,W]={\rm i}\theta{\bf 1}$. Some suitable completion of $U({\cal H})$, denoted $\overline{U({\cal H})}$ and to be constructed presently, is the space of convergent power series in $V,W$. We take an arbitrary vector of $\overline{U({\cal H})}$ to be an expression of the form
\begin{equation}
\psi(V,W)=\sum_{n,m=0}^{\infty}\frac{c_{nm}}{\sqrt{n!m!\,\theta^{n+m}}}V^nW^m,
\label{funzia}
\end{equation}
where the $c_{nm}$ are complex coefficients, such that the above series converges (in a sense to be specified presently). The factor $(\theta^{n+m})^{-1/2}$ ensures that all summands are dimensionless.  From now we will prescribe all vectors of $\overline{U({\cal H})}$  to be normal--ordered, {\it i.e.}, $V$ will {\it always}\/ be assumed to precede $W$, if necessary by applying the commutator $[V,W]={\rm i}\theta{\bf 1}$.

A basis for $\overline{U({\cal H})}$ is given by the vectors
\begin{equation}
\psi_{nm}(V,W)= \frac{1}{\sqrt{n!m!\,\theta^{n+m}}}V^{n}W^{m}, \qquad  n,m\in\mathbb{N}.
\label{sonrio}
\end{equation}
The simplest choice for a scalar product on $\overline{U({\cal H})}$ is to declare the basis vectors (\ref{sonrio}) orthonormal,
\begin{equation}
\langle\psi_{n_1m_1}\vert\psi_{n_2m_2}\rangle:=\delta_{n_1n_2}\delta_{m_1m_2},
\label{skalar}
\end{equation}
and to extend (\ref{skalar}) to all of $\overline{U({\cal H})}$ by complex linearity. Then the squared norm of the vector (\ref{funzia}) equals $\sum_{nm}\vert c_{nm}\vert^2$:
\begin{equation}
\vert\vert\psi(V,W)\vert\vert^2=\sum_{n,m=0}^{\infty}\vert c_{nm}\vert^2.
\label{norma}
\end{equation}
Since this norm must be finite, this identifies $\overline{U({\cal H})}$ as the Hilbert space of square--summable complex sequences $\left\{c_{nm}\right\}$ in two indices $n,m$, {\it the latter taken to be normal--ordered}\/ as in (\ref{sonrio}); this defines the completion of $U({\cal H})$ referred to above.
It is worthwhile to observe that, although the vectors (\ref{funzia}) are unbounded operators in their action on the auxiliary Hilbert space $L^2(\mathbb{R}, {\rm d}v)$, the same vectors {\it do}\/ have a finite norm as elements of the Hilbert space $\overline{U({\cal H})}$. This is so because the norm of $\psi(V,W)$ in (\ref{norma}) is being measured by means of the complex coefficients $c_{nm}$, not by means of the operator norms of $V,W$ (themselves infinite). We will henceforth call the $\psi_{nm}$ of (\ref{sonrio}) {\it noncommutative oscillator modes}\/.

The Hilbert space $\overline{U({\cal H})}$ just constructed will become the carrier space of a representation of the algebra  (\ref{possonalg}). For this we need to define the action of the operators $X,Y,P_X,P_Y$ on the noncommutative oscillator modes (\ref{sonrio}). We set
\begin{equation}
X\psi_{nm}:=\sqrt{\frac{\theta}{2}}\left(\sqrt{n+1}\,\psi_{n+1,m}+\sqrt{n}\,\psi_{n-1,m}\right)
\label{equis}
\end{equation}
and
\begin{equation}
P_X\psi_{nm}:=\frac{{\rm i}\hbar}{\sqrt{2\theta}}\left(\sqrt{n+1}\,\psi_{n+1,m}-\sqrt{n}\,\psi_{n-1,m}\right).
\label{pequis}
\end{equation}
{}For the second index we define the action of $Y,P_Y$ similarly, with the sole difference that the (reverse) Bopp shift (\ref{vop}) must be taken into account:
\begin{equation}
Y\psi_{nm}:=\sqrt{\frac{\theta}{2}}\left(\sqrt{m+1}\,\psi_{n,m+1}+\sqrt{m}\,\psi_{n,m-1}\right)+\frac{\theta}{\hbar}P_X\psi_{nm}
\label{ygriega}
\end{equation}
and
\begin{equation}
P_Y\psi_{nm}:=\frac{{\rm i}\hbar}{\sqrt{2\theta}}\left(\sqrt{m+1}\,\psi_{n,m+1}-\sqrt{m}\,\psi_{n,m-1}\right).
\label{pygriega}
\end{equation}
{}Finally, the operators $X,Y,P_X,P_Y$ so defined are Hermitian and  satisfy the algebra (\ref{possonalg}) as desired. The above $X,Y,P_X,P_Y$ are distinguished notationally from the operators $X',Y',P'_X,P'_Y$ of (\ref{unoequis})--(\ref{unopygriega}) in order to stress the fact that they are actually different operators acting on different spaces\footnote{All infinite--dimensional, complex, separable Hilbert spaces being unitarily isomorphic, the above statement is to be understood as {\it different realisations of Hilbert space}.}, even if the two sets of operators satisfy the same algebra (\ref{possonalg}). From now on we will only work with the representation of the algebra (\ref{possonalg}) provided by (\ref{equis})--(\ref{pygriega}).

Although they will not be used here, the previous results can be easily generalised to higher dimensions \cite{EIKONALMODES}.

\section{The Hamilton--Jacobi equation on the Moyal plane}\label{hjmp}

Our next task is to write down the Hamilton--Jacobi equation. For this we define the following dimensionless coordinates $Q_A, Q_B$ and momenta $P_A,P_B$:
\begin{equation}
Q_A:=\frac{1}{\sqrt{\theta}}X,\quad P_A:=\frac{\sqrt{\theta}}{\hbar}P_X, \quad Q_B:=\frac{1}{\sqrt{\theta}}Y-\frac{\sqrt{\theta}}{\hbar}P_X, \quad P_B:=\frac{\sqrt{\theta}}{\hbar}P_Y.
\label{rml}
\end{equation}
These operators satisfy the standard, dimensionless,  Poisson--Heisenberg algebra:
\begin{equation}
[Q_A,P_A]=[Q_B,P_B]={\rm i}{\bf 1},\quad [Q_A,Q_B]=[P_A,P_B]=[Q_A,P_B]=[Q_B,P_A]=0.
\label{identik}
\end{equation}
One can think of the space spanned by $Q_A,Q_B,P_A,P_B$ as a commutative phase space, the only difference being that coordinates and momenta are operators on $\overline{U({\cal H})}$. Correspondingly, phase--space derivatives will be replaced with the adjoint action (\ref{adjk}). Our strategy will be to first write down the Hamilton--Jacobi equation on this commutative phase space. Then we will transform the result back into the noncommutative space spanned by $X,Y,P_X,P_Y$.

A key property of the classical mechanical action ${\cal S}$, when expressed as a function of the coordinates as in eqn. (\ref{aktion}), is that it serves as a potential function for the momenta, {\it i.e.}, $p_x=\partial {\cal S}/\partial x$ and $p_y=\partial {\cal S}/\partial y$. This property must be maintained in the case under consideration here, where coordinates and momenta are operator--valued, and the adjoint action replaces the partial derivatives. Thus we need to find a Hermitian operator, that we will call the {\it operator--valued action}\/ $S$, depending on $Q_A,P_A,Q_B,P_B$, and such that it will yield the momenta when one takes the adjoint action with respect to the coordinates. In order to obtain a linear expression in the momenta, we need $S$ to be a quadratic combination of the momenta. This leads one to the following operator:
\begin{equation}
S:=-\frac{1}{\hbar}Et{\bf 1}+\frac{1}{2}P_A^2+\frac{1}{2}P_B^2-U(Q_A,Q_B).
\label{sepa}
\end{equation}
Here $U(Q_A,Q_B)$ is a dimensionless real function of $Q_A,Q_B$, that we can look upon as an operator--valued generalisation of the classical potential function ${\cal U}(x,y)$ of eqn. (\ref{kaspos}). Indeed, whatever our choice for $U(Q_A,Q_B)$ we find
\begin{equation}
{\rm i}P_A={\rm ad}_{Q_A}\left(S\right), \qquad {\rm i}P_B={\rm ad}_{Q_B}\left(S\right)
\label{rpll}
\end{equation}
as one should; the factors of $i$ ensure the Hermitian property. Eqns. (\ref{sepa}), (\ref{rpll}) are to be regarded as the noncommutative generalisation of eqn. (\ref{aktion}).  We would like to observe that the following consistency check on (\ref{rpll}) is satisfied. The integrability condition $\partial^2{\cal S}/\partial y\partial x= \partial p_x/\partial y=\partial p_y/\partial x=\partial^2{\cal S}/\partial x\partial y$ holds true in eqn. (\ref{aktion}). Therefore the operator analogue of  this classical integrability condition should read
\begin{equation}
{\rm ad}_{Q_A}(P_B)={\rm ad}_{Q_B}(P_A),
\label{hwak}
\end{equation}
and, indeed, this is satisfied thanks to the Jacobi identity (\ref{homor}).

The operator action $S$ is a dimensionless, Hermitian quantum operator acting on the carrier space $\overline{U({\cal H})}$. Now, in order to write down the Hamilton--Jacobi equation, a Hamiltonian is needed. We will make a judicious choice for the Hamiltonian operator, followed by some consistency checks to ensure that our choice is correct. We claim that the Hamiltonian operator $H$ correponding to (\ref{sepa}) is given by
\begin{equation}
H=\frac{1}{2}P_A^2+\frac{1}{2}P_B^2+U(Q_A,Q_B).
\label{jamil}
\end{equation}
The above is also a dimensionless, Hermitian operator. Replacing phase--space derivatives with adjoint actions, it is reasonable to demand that the Hamilton equations of motion be
\begin{equation}
\dot P_A=-{\rm ad}_{Q_A}(H),\quad \dot Q_A={\rm ad}_{P_A}(H)\quad \dot P_B=-{\rm ad}_{Q_B}(H),\quad \dot Q_B={\rm ad}_{P_B}(H).
\label{kanonik}
\end{equation}
We find, for the Hamiltonian (\ref{jamil}) and the canonical pair $Q_A,P_A$,
\begin{equation}
{\rm ad}_{P_A}(H)=-{\rm i}\frac{\partial U}{\partial Q_A}, \qquad {\rm ad}_{Q_A}(H)={\rm i}P_A.
\label{judic}
\end{equation}
Thus Newton's law is satisfied as it should, because
\begin{equation}
\ddot Q_A=\frac{{\rm d}}{{\rm d}t}\left({\rm ad}_{P_A}(H)\right)={\rm ad}_{\dot {P_{A}}}(H)=
-[[Q_A,H],H]=-{\rm i}[P_A,H]=-\frac{\partial U}{\partial Q_A}.
\label{niuto}
\end{equation}
Obviously the same holds for the other canonical pair $Q_B,P_B$.

We can now write down the noncommutative Hamilton--Jacobi equation for a particle of mass $m$ on the Moyal plane, subject to the potential $U(Q_A,Q_B)$. It reads
\begin{equation}
\frac{\partial S}{\partial t}+\frac{\hbar}{m\theta}\left[-\frac{1}{2}\left({\rm ad}_{Q_A}(S)\right)^2-\frac{1}{2}\left({\rm ad}_{Q_B}(S)\right)^2+U(Q_A,Q_B)\right]=0.
\label{pld}
\end{equation}
We draw attention to the negative sign preceding the squared adjoint actions, due to the imaginary units in (\ref{rpll}); otherwise (\ref{pld}) is the natural operator generalisation of its classical counterpart (\ref{kaspos}). The factor $\hbar/(m\theta)$ has the dimensions of time inverse, thus making (\ref{pld}) dimensionally homogeneous. We will find it useful to separate out in (\ref{sepa}) the piece that is proportional to the identity, thus leaving the reduced, or time--independent, operator action $S^{(0)}$:
\begin{equation}
S=-\frac{1}{\hbar}Et{\bf 1}+S^{(0)}, \quad S^{(0)}:=\frac{1}{2}P_A^2+\frac{1}{2}P_B^2-U(Q_A,Q_B).
\label{zeit}
\end{equation}
Then (\ref{rpll}) becomes
\begin{equation}
{\rm i}P_A={\rm ad}_{Q_A}\left(S^{(0)}\right), \qquad {\rm i}P_B={\rm ad}_{Q_B}\left(S^{(0)}\right),
\label{ripoll}
\end{equation}
which gives the time--independent Hamilton--Jacobi equation
\begin{equation}
\frac{\hbar^2}{m\theta}\left[-\frac{1}{2}\left({\rm ad}_{Q_A}(S^{(0)})\right)^2-\frac{1}{2}\left({\rm ad}_{Q_B}(S^{(0)})\right)^2+U(Q_A,Q_B)\right]=E.
\label{bbmk}
\end{equation}
Here appears the quantity $\hbar^2/(m\theta)$ mentioned in section \ref{nonpo}.

A comment is in order. In principle one would not expect Planck's constant $\hbar$ to be present in the Hamilton--Jacobi equation, since the latter is a classical equation, which arises before quantisation. This much is true of theories on commutative spaces. However, as remarked in section \ref{einfuehrung}, any theory on noncommutative space must include $\hbar$ because, on an energy scale, quantum effects set in much earlier than noncommutative effects. This being the case, the distinction between {\it classical}\/ and {\it quantum}\/ turns out to be rather formal.

A more mundane explanation of the same fact is provided by the following argument. The noncommutative theory depends on the dimensionful parameter $\theta$. The latter must enter the Hamilton--Jacobi equation. Now (\ref{pld}) and (\ref{bbmk}) cannot be balanced dimensionally in terms of just one dimensionful parameter; at least one more dimensionful parameter is needed for homogeneity. Planck's constant $\hbar$ does precisely that job.

Using (\ref{rml}) we can now rewrite the operator action of $(\ref{sepa})$ in terms of $X,Y,P_X,P_Y$:
\begin{equation}
S:=-\frac{1}{\hbar}Et{\bf 1}+\frac{\theta}{2\hbar^2}P_X^2+\frac{\theta}{2\hbar^2}P_Y^2-U(X,Y,P_X).
\label{kkmd}
\end{equation}
Some caution is necessary here since, in general, the potential function $U(Q_A, Q_B)$ suffers from ordering ambiguities once we express $Q_A,Q_B$ in terms of $X,Y,P_X,P_Y$. This requires that some ordering prescription be adopted, {\it e.g.}, Weyl's symmetrisation\footnote{This is not specific to our approach in terms of operator--valued quantities, since the same ordering ambiguities would arise if we used a star product.}. We also observe that the potential $U$ in (\ref{kkmd}) can depend on $P_X$, but not on $P_Y$, due to the Bopp shift (\ref{vop}). From the time--independent operator action $S^{(0)}$ of (\ref{zeit}) we similarly obtain
\begin{equation}
S^{(0)}:=\frac{\theta}{2\hbar^2}P_X^2+\frac{\theta}{2\hbar^2}P_Y^2-U(X,Y,P_X).
\label{bbhjp}
\end{equation}
{}For the time--dependent Hamilton--Jacobi equation (\ref{pld}) we find
\begin{equation}
\frac{\partial S}{\partial t}+\frac{\hbar}{m\theta}\left[-\frac{1}{2\theta}\left({\rm ad}_{X}(S)\right)^2-\frac{1}{2\theta}\left({\rm ad}_{Y}(S)-\frac{\theta}{\hbar}{\rm ad}_{P_X}(S)\right)^2+U\right]=0,
\label{rjs}
\end{equation}
while its time--independent version (\ref{bbmk}) becomes
\begin{equation}
\frac{\hbar^2}{m\theta}\left[-\frac{1}{2\theta}\left({\rm ad}_{X}(S^{(0)})\right)^2-\frac{1}{2\theta}\left({\rm ad}_{Y}(S^{(0)})-\frac{\theta}{\hbar}{\rm ad}_{P_X}(S^{(0)})\right)^2+U\right]=E.
\label{bbmd}
\end{equation}

Altogether, eqns. (\ref{rjs}) and (\ref{bbmd}) above reexpress the Hamilton--Jacobi equations (\ref{pld}) and (\ref{bbmk}) in terms of the noncommutative variables $X,Y,P_X,P_Y$. However, in general one should stop short of calling  (\ref{rjs}) and (\ref{bbmd}) Hamilton--Jacobi equations in the strict sense of the word. For such to be the case, one should be able to replace any possible occurrence of $P_X$ with its expression in terms of ${\rm ad}_X(S)$. One such occurrence happens within the potential $U$. This makes the replacement impossible, as we see from (\ref{kkmd}), because one has $P_X=\hbar \,{\rm ad}_X(S+U)/({\rm i}\theta)$: in trying to eliminate $P_X$ in favour of ${\rm ad}_X(S)$, the offending term in the potential $U$ reappears! Moreover, $P_X$ also shows up in the terms ${\rm ad}_{P_X}(S)$ and ${\rm ad}_{P_X}(S^{(0)})$, where it should also be replaced.

A moment's reflection shows that, in fact, things are exactly as they should. Let us go back to eqns. (\ref{aktion}), (\ref{kaspos}), where it is implicitly understood that $x,y,p_x,p_y$ satisfy the standard Poisson algebra $\{x,y\}=0=\{p_x,p_y\}$, $\{x,p_y\}=0=\{y,p_x\}$, $\{x,p_x\}=1=\{y,p_y\}$, which is isomorphic to that in (\ref{identik}). All these variables are canonical. This fact guarantees that the replacements $p_x=\partial {\cal S}/\partial x$ and $p_y=\partial {\cal S}/\partial y$, as well as their operator--valued analogues ${\rm i}P_A={\rm ad}_{Q_A}(S)$, ${\rm i}P_B={\rm ad}_{Q_B}(S)$, can be performed. Thus (\ref{pld}) and (\ref{bbmk}) are {\it bona fide}\/ Hamilton--Jacobi equations. However, neither the Bopp shift (\ref{vop}) nor its inverse is a canonical transformation, because the algebra satisfied by $X,Y,P_X,P_Y$ differs from that satified by $Q_A,Q_B,P_A,P_B$. The latter are canonical variables, while the former are not.

To summarise, we have written down the Hamilton--Jacobi equation using a set of (operator--valued) canonical variables $Q_A,Q_B,P_A,P_B$, and we have then transformed the resulting equation using a set of noncanonical variables $X,Y,P_X,P_Y$, by means of a diffeomorphism (the Bopp shift) that does {\it not}\/ qualify as a canonical transformation. There is no way the Moyal phase space $\mathbb{R}^4_{\theta, \hbar}$ can be {\it canonically}\/ transformed into the standard phase space $\mathbb{R}^4_{\hbar}$. Physically this is so because the quantum of area $\theta$ that is present in $\mathbb{R}^4_{\theta, \hbar}$ is absent in $\mathbb{R}^4_{\hbar}$. The Bopp shift respects the quantum of action $\hbar$, but not the quantum of area $\theta$.

\section{The Schroedinger equation on the Moyal plane}\label{plkdmr}

In order to write down the Schroedinger equation on the Moyal plane, we will follow the same strategy of section \ref{hjmp}. Namely, we will first work with the canonical variables $Q_A,Q_B,P_A,P_B$  of (\ref{rml}), in terms of which we will write down the Schroedinger equation; only then will we transform back to the noncommutative variables $X,Y,P_X,P_Y$.

The Schroedinger equation we will arrive at will turn out to be valid only semiclassically. We first need explain what one understands as the semiclassical limit of noncommutative quantum mechanics. In the commutative case, the semiclassical limit is obtained as $\hbar\to 0$, when the Schroedinger equation reduces to the Hamilton--Jacobi equation. Since noncommutative quantum mechanics contains two deformation parameters $\hbar, \theta$, we ask what the precise regime of these parameters is that corresponds to the eikonal approximation. We claim that the eikonal approximation corresponds to the limit $\hbar\to 0$ and $\theta\to 0$ while holding $\hbar^2/(m\theta)$ fixed. Obviously $\hbar$ must go to zero. However, as mentioned in the introduction, noncommutative effects set in (on an energy scale) much later than quantum effects, so $\hbar\to 0$ enforces $\theta\to 0$ as well. Since the ratio $\hbar^2/(m\theta)$ must be held fixed for the Hamilton--Jacobi equation (\ref{pld}) (or its reexpression (\ref{rjs})) to be well defined, this proves our claim.

To begin with, let us consider the free case, $U=0$. We expect a time--independent, semiclassical wavefunction $\Phi^{(0)}$ to be given by the exponential of ($i$ times) the reduced action of eqn. (\ref{zeit}):
\begin{equation}
\Phi^{(0)}=\exp\left({\rm i}S^{(0)}\right)=\exp\left(\frac{{\rm i}}{2}P_A^2+\frac{{\rm i}}{2}P_B^2\right).
\label{ttenn}
\end{equation}
Using the algebra (\ref{identik}) we find
\begin{equation}
{\rm ad}_{Q_A}\Phi^{(0)}=-P_A\Phi^{(0)}, \qquad
{\rm ad}_{Q_B}\Phi^{(0)}=-P_B\Phi^{(0)}
\label{papea}
\end{equation}
and
\begin{equation}
{\rm ad}_{Q_A}^2\Phi^{(0)}=\left(P_A^2-{\rm i}{\bf 1}\right)\Phi^{(0)}, \quad
{\rm ad}_{Q_B}^2\Phi^{(0)}=\left(P_B^2-{\rm i}{\bf 1}\right)\Phi^{(0)}.
\label{papeb}
\end{equation}
Remembering (\ref{ripoll}) we arrive at
\begin{equation}
\frac{1}{2}\left({\rm ad}_{Q_A}^2+{\rm ad}_{Q_B}^2+2{\rm i}\right)\Phi^{(0)}
=-\frac{1}{2}\left[\left({\rm ad}_{Q_A}(S^{(0)})\right)^2+\left({\rm ad}_{Q_B}(S^{(0)})\right)^2\right]\Phi^{(0)}.
\label{kompp}
\end{equation}
Now eqn.  (\ref{bbmk}) suggests equating the right--hand side to $Em\theta\Phi^{(0)}/\hbar^2$:
\begin{equation}
\frac{1}{2}\left({\rm ad}_{Q_A}^2+{\rm ad}_{Q_B}^2+2{\rm i}\right)\Phi^{(0)}
=\frac{Em\theta}{\hbar^2}\Phi^{(0)}.
\label{eliff}
\end{equation}
Setting $\Phi:=\Phi^{(0)}\exp\left(-{\rm i}Et/\hbar\right)$ we can finally write
\begin{equation}
\frac{\hbar^2}{2m\theta}\left({\rm ad}_{Q_A}^2+{\rm ad}_{Q_B}^2+2{\rm i}\right)\Phi
={\rm i}{\hbar}\frac{\partial\Phi}{\partial t}.
\label{eliffee}
\end{equation}

Let us take stock. The expression ${\rm ad}_{Q_A}^2+{\rm ad}_{Q_B}^2$ on the left--hand side can be interpreted as an operator--valued analogue of the standard Laplacian $\partial^2/\partial x^2 + \partial^2/\partial x^2$. The term $2{\rm i}{\bf 1}$ can be interpreted as a constant potential, and can therefore be dropped. As it stands, (\ref{eliffee}) is strictly equivalent to the Hamilton--Jacobi equation (\ref{pld}) when $U=0$, and we can declare
\begin{equation}
\frac{\hbar^2}{2m\theta}\left({\rm ad}_{Q_A}^2+{\rm ad}_{Q_B}^2\right)\Phi
={\rm i}{\hbar}\frac{\partial\Phi}{\partial t}
\label{eliffeeww}
\end{equation}
to be the Schroedinger equation for a free particle on the Moyal plane. Modulo the factor $\hbar^2/2m\theta$, eqn. (\ref{eliffeeww}) is formally identical to the standard Schrodinger equation. However it must be borne in mind that its structure is substantially different. Eqn. (\ref{eliffeeww}) is {\it not}\/ the expression of an operator acting on a vector, to produce another vector. Rather, it expresses an equality between operators. By the same token, its time--independent form (\ref{eliff}) is not a eigenvalue equation for a vector, but an eigenvalue equation for the eigen{\it operator}\/ $\Phi^{(0)}$. (This is the operator analogue of the star--eigenvalue equations; see, {\it e.g.}, ref. \cite{ZACHOS}). Last but not least, we recall that no approximation has been made in order to reproduce the Hamilton--Jacobi equation from the Schroedinger equation, as (\ref{eliffeeww}) and (\ref{pld}) are strictly equivalent when $U=0$. We will see presently that this equivalence will also remain in the interacting case, at least in the semiclassical limit.

In the presence of a potential $U$, the natural generalisation of (\ref{eliffeeww}) is
\begin{equation}
\frac{\hbar^2}{2m\theta}\left[{\rm ad}_{Q_A}^2+{\rm ad}_{Q_B}^2+U(Q_A,Q_B)\right]\Psi
={\rm i}{\hbar}\frac{\partial\Psi}{\partial t}.
\label{qqeliffeeww}
\end{equation}
We look for semiclassical solutions to (\ref{qqeliffeeww}) in the form $\Psi:=\Psi^{(0)}\exp\left(-{\rm i}Et/\hbar\right)$, where $\Psi^{(0)}$ is suggested by (\ref{zeit}):
\begin{equation}
\Psi^{(0)}:=\exp\left({\rm i}S^{(0)}\right)=\exp\left[\frac{{\rm i}}{2}P_A^2+\frac{{\rm i}}{2}P_B^2-{\rm i}U(Q_A,Q_B)\right].
\label{losung}
\end{equation}
Unfortunately there is no neat expression for the analogues of (\ref{papea}) and (\ref{papeb}) when $U$ is nontrivial.  One can power--expand the exponential (\ref{losung}) and act with ${\rm ad}_{Q_A}$,  ${\rm ad}_{Q_B}$ term by term, but the presence of a nonconstant $U(Q_A,Q_B)$ prevents a tidy rearrangement of the result into any manageable expression. This is ultimately due to the fact that, when $U$ is nonconstant,  (\ref{losung}) does {\it not}\/ factorise as
\begin{equation}
\exp\left(\frac{{\rm i}}{2}P_A^2+\frac{{\rm i}}{2}P_B^2\right)\exp\left[-{\rm i}U(Q_A,Q_B)\right].
\label{implicc}
\end{equation}
In turn, the impossibility of the factorisation (\ref{implicc}) is due to the nonvanishing of the following commutators:
\begin{equation}
[P_A^2, U]=-{\rm i}P_A\frac{\partial U}{\partial Q_A}-{\rm i}\frac{\partial U}{\partial Q_A}P_A,\qquad
[P_B^2, U]=-{\rm i}P_B\frac{\partial U}{\partial Q_B}-{\rm i}\frac{\partial U}{\partial Q_B}P_B.
\label{fatto}
\end{equation}
However, we should remember that the commutators (\ref{fatto}) have been computed using the dimensionless algebra (\ref{identik}). When one reinstates powers of $\hbar$, one immediately sees that the right--hand sides of (\ref{fatto}) are $O(\hbar)$. In the semiclassical limit considered throughout in this paper, one may drop terms of order $\hbar$ while keeping $\hbar^2/(m\theta)$ fixed. We may thus approximate the right--hand sides of (\ref{fatto}) by zero. In this limit, the wavefunction (\ref{losung}) can be approximated by its factorised form (\ref{implicc}):
\begin{equation}
\Psi^{(0)}\simeq\exp\left(\frac{{\rm i}}{2}P_A^2+\frac{{\rm i}}{2}P_B^2\right)\exp\left[-{\rm i}U(Q_A,Q_B)\right].
\label{impqqlicc}
\end{equation}
Using the semiclassical wavefunction (\ref{impqqlicc}), one sees that the reasoning from eqn. (\ref{papea}) to eqn. (\ref{eliffeeww}) continues to hold true in the presence of the potential $U(Q_A,Q_B)$. In this way one establishes that the operator wavefunction $\Psi:=\Psi^{(0)}\exp\left(-{\rm i}Et/\hbar\right)$ satisfies the Schroedinger equation (\ref{qqeliffeeww}).  For the latter we claim validity within the semiclassical regime only, given the approximation made in (\ref{impqqlicc}). Moreover, as was already the case for the free particle, the Schroedinger equation (\ref{qqeliffeeww}) in the presence of a potential $U$ is strictly equivalent to the Hamilton--Jacobi equation (\ref{pld}). In this sense, the difference between these two equations lies in the choice one makes for the quantity one works with, {\it i.e.}, either the action $S$ or its exponential. We should also add that the reverse order for the factors in (\ref{impqqlicc}) would be justified just as well in the semiclassical limit. Within the accuracy of this limit, it is actually a matter of choice which exponential appears on the left and which one on the right.

As a final step, we need to recast the Schroedinger equation (\ref{qqeliffeeww}) in terms of the noncommutative variables $X,Y,P_X,P_Y$. This is readily done:
using (\ref{rml}) we perform the replacement
\begin{equation}
{\rm ad}^2_{Q_A}+{\rm ad}^2_{Q_B}=\frac{1}{\theta}{\rm ad}^2_{X}+\frac{1}{\theta}{\rm ad}^2_Y-\frac{2}{\hbar}{\rm ad}_{P_X}{\rm ad}_{Y}
+\frac{\theta}{\hbar^2}{\rm ad}^2_{P_X}
\label{reemp}
\end{equation}
in (\ref{qqeliffeeww}). This gives
\begin{equation}
\frac{\hbar^2}{2m\theta}\left[\frac{1}{\theta}{\rm ad}^2_{X}+\frac{1}{\theta}{\rm ad}^2_Y-\frac{2}{\hbar}{\rm ad}_{P_X}{\rm ad}_{Y}
+\frac{\theta}{\hbar^2}{\rm ad}^2_{P_X}+U(X,Y,P_X)\right]\Psi
={\rm i}{\hbar}\frac{\partial\Psi}{\partial t}.
\label{qqell}
\end{equation}
The same {\it caveat}\/ discussed at length after eqn. (\ref{bbmd}) applies to (\ref{qqell}) as well.

\section{Discussion}\label{diskku}

In this paper we have taken some first steps towards a head--on approach to quantum mechanics on noncommutative spaces, an approach that has been demanded and studied to some extent in the literature \cite{ZA4}. The novelty lies in the attempt to express wavefunctions purely in terms of operator--valued coordinates, rather than in terms of c--valued functions that are multiplied together by means of a star product.  The underlying logic is as follows. Coordinates $X,Y$ on the Moyal plane are operators satisfying $[X,Y]={\rm i}\theta{\bf 1}$. This implies that wavefunctions $\Psi$, as functions of $X,Y$, must also be operators. This represents a radical departure from the viewpoint of deformation quantisation, where noncommutativity lies hidden under the star product of c--valued wavefunctions. Not only wavefunctions, but the mechanical action itself (the solution to the Hamilton--Jacobi equation) must become an operator. This is totally natural since, at least in the semiclassical limit, one expects the mechanical action to be proportional to the logarithm of the wavefunction. If the latter is an operator, so must be the former.

The strategy followed in writing down the Hamilton--Jacobi and the  Schroedinger equations on the Moyal plane involves three steps. The first step is to use the Bopp shift (\ref{vop}), in order to transform the original noncommutative variables $X,Y,P_X,P_Y$ (satisfying the algebra (\ref{possonalg})) into dimensionless variables $Q_A,Q_B,P_A,P_B$ (satisfying the algebra (\ref{identik})). In terms of the latter there is a well--defined procedure for writing down the Hamilton--Jacobi equation. The second step is to pass therefrom to the Schroedinger equation. This second step involves some generally accepted guesswork\footnote{This guesswork is sometimes summarised in the statement that {\it first quantisation is a mystery, second quantisation is a functor}.}. The third, and final, step, is to undo the Bopp shift and transform the equations so obtained back into the original noncommutative variables $X,Y,P_X, P_Y$. The Bopp shift is a diffeomorphism that does {\it not}\/ qualify as a canonical transformation. However we need canonical variables in order to first write down the Hamilton--Jacobi equation, which one later uses as a bridge to the Schroedinger equation.  As explained in detail towards the end of section \ref{hjmp}, it is impossible to {\it canonically}\/ transform the Moyal phase space $\mathbb{R}^4_{\theta, \hbar}$ into the standard phase space $\mathbb{R}^4_{\hbar}$. Physically this is so because the quantum of area $\theta$ that is present in $\mathbb{R}^4_{\theta, \hbar}$ is absent in $\mathbb{R}^4_{\hbar}$. The existence of the two quanta $\hbar$ and $\theta$ leads to the existence of a natural energy scale  $\hbar^2/(m\theta)$ (for any given particle mass $m$), which is absent in standard quantum mechanics.

A key element in our construction is provided by the noncommutative oscillator modes $\psi_{nm}$ of section \ref{sstres}. The $\psi_{nm}$ are quantum mechanical wavefunctions of a harmonic oscillator defined on (an auxilary copy of) the Moyal plane. As $\theta\to 0$, the $\psi_{nm}$ must be replaced with the commutative oscillator modes $\phi_{nm}$ of section \ref{com} (the $\phi_{nm}$ are standard oscillator modes on $\mathbb{R}^2$). Finally setting $\theta=0$ but $\sqrt{\theta}=1$ (as befits the fact that $\sqrt{\theta}>\theta$ when $\theta\to 0$) we see that eqns. (\ref{equis}) to (\ref{pygriega}) respectively become eqns. (\ref{unoequis}) to (\ref{unopygriega}): this is the commutative limit.

The symmetry algebra (the commutator algebra of section \ref{ssuno}) is realised unitarily on the Hilbert space $\overline{U({\cal H})}$ spanned by the noncommutative oscillator modes $\psi_{nm}$. The latter are not to be confused with the true quantum states $\Psi$ of the theory. We thus meet a situation in which the quantum states $\Psi$ of the theory do {\it not}\/ support a representation of the symmetry algebra---in apparent violation of Wigner's theorem. There is however no violation, because Wigner's theorem implicitly assumes a commutative space. The states $\psi_{nm}$ that support a representation of the symmetry algebra are intermediate states in our construction, while the true quantum states $\Psi$, being operator--valued and thus noncommutative, are not bound by Wigner's theorem to furnish a representation. Similar arguments apply to the Stone--von Neumann theorem as applied to the subalgebra $[X,P_X]={\rm i}\hbar=[Y,P_Y]$. This latter theorem is also not violated since it too presupposes a commutative space.

The following thoughts, of a somewhat speculative nature, are collected here to conclude. It was mentioned in the introduction, and also right after the Hamilton--Jacobi equation (\ref{bbmk}) that, in the presence of noncommutativity, the distinction between {\it classical}\/ and {\it quantum}\/  turns out to be somewhat formal, devoid of physical content. This is so because, in principle, one does not expect Planck's constant to arise at the level of the classical Hamilton--Jacobi equation---but the fact is, it does arise. There is also no way one can have a {\it purely classical}\/ noncommutative theory because quantum effects set in much earlier, on an energy scale, than noncommutative effects. Moreover, when the potential is constant on the Moyal plane, the Schroedinger equation (for the exponential of the action operator) and the Hamilton--Jacobi equation (for the action operator alone) are actually equivalent. This is in marked contrast with the case of commutative quantum mechanics, where the same equivalence holds only semiclassically.  In the interacting case on the Moyal plane this equivalence (between Schroedinger and Hamilton--Jacobi) is generally lost (of course, it continues to hold in the semiclassical limit).  One is thus tempted to call this state of affairs a {\it classical/quantum duality}\/ of noncommutative quantum mechanics. It is interesting to observe that analogous effects have been reported in \cite{SINGH, CARROLL0, ME}. Although the latter refer to somewhat different contexts, they are by no means totally different from ours. One is also reminded of the UV/IR mixing of noncommutative field theories \cite{SZABO2}. Altogether, we find these similarities very suggestive.

\vskip.5cm
\noindent
{\bf Acknowledgements}    J.M.I. thanks Max--Planck--Institut f\"ur Gravitationsphysik, Albert--Einstein--Institut (Golm, Germany), for hospitality. This work has been supported by Universidad Polit\'ecnica de Valencia under grant PAID-06-09, and by Generalitat Valenciana (Spain).

\end{document}